\documentclass[reprint,prb,twocolumn,showpacs,superscriptaddress,aps]{revtex4-1}

\usepackage{graphicx}
\DeclareGraphicsExtensions{.png,.jpg,.eps}

\usepackage{xcolor}
\usepackage{amsmath}
\usepackage{caption}
\usepackage{subcaption}
\usepackage{float}

\usepackage{wrapfig}

\begin{document}

\title{Ab initio study of the strain dependence of thermopower in electron-doped SrTiO$_3$}

\author{Adolfo O. Fumega}
  \email{adolfo.otero@rai.usc.es}
\author{V. Pardo}
  \email{victor.pardo@usc.es}
\affiliation{Departamento de F\'{i}sica Aplicada,
  Universidade de Santiago de Compostela, E-15782 Campus Sur s/n,
  Santiago de Compostela, Spain}
\affiliation{Instituto de Investigaci\'{o}ns Tecnol\'{o}xicas,
  Universidade de Santiago de Compostela, E-15782 Campus Sur s/n,
  Santiago de Compostela, Spain}


\begin{abstract} 

In this paper we explore the different mechanisms that affect the thermopower of a band insulating perovskite (in this case, SrTiO$_3$) when subject to strain (both compressive 
or tensile). We analyze the high temperature, entropy dominated limit and the lower temperature, energy-transport regime. We observe that the effect of strain in the 
high-temperature Seebeck coefficient is small at the concentration levels of interest for thermoelectric applications. However, the effective mass changes substantially with strain, 
which produces an opposite effect to that of the degeneracy-breakups brought about by strain. In particular, we find that the thermopower can be enhanced by applying tensile strain in the adequate regime. We conclude that the detrimental effect of strain in thermopower due to 
band splitting is a minor effect that will not hamper the optimization of the thermoelectric properties of oxides with t$_{2g}$-active bands by applying strain.

\end{abstract}
\maketitle

\section{Introduction}

Oxides have become an interesting playground to search for new materials with enhanced thermoelectric (TE) properties. They can be easily synthesized, they are chemically stable, they can be nanostructured in the form of films,\cite{oxide_films} nanoparticles\cite{manganite_nanoparticles} and nanowires,\cite{tom_wu_vo2_nanowires} and they span a huge range of chemical compositions and doping.\cite{oxide_electronics_mannhart} Examples of oxides with promising TE properties are the family of misfit layered cobaltates\cite{hebert,Terasaki_cobalt} and also SrTiO$_3$\cite{sto_TE} and its nanostructures.\cite{sto_2deg_TE}

In standard semiconductors, the parameters entering the TE figure of merit,\cite{snyder2008} namely the electrical conductivity, the Seebeck coefficient and the thermal conductivity (in particular its electronic component) are usually coupled so that improving the electrical conductivity has a negative effect in the Seebeck coefficient and the thermal conductivity, and vice versa. Many different routes have been proposed to optimize the figure of merit.\cite{lu_high_figure_merit_2016,zheng_improvement_2013} One way to enhance the TE response of a material is, on one hand, to play with the phonon part of the thermal conductivity,\cite{yu_thermal_2008,zT_phonon} and in order to decouple the electronic components, strain has been suggested as a promising mechanism to do so.\cite{zT_strain_sp} Oxides, with their flexible d electrons, should be more permeable to changes and tunability of their electronic structure with strain\cite{la2nio4_strain} and nanostructuring\cite{zT_oxides_nano} than standard sp semiconductors. d electron systems show many different phases as a product of the coupling between lattice, spin, charge and orbital degrees of freedom,\cite{khomskii_review_dof} which could help tuning the band structure at will and modify the TE response in the desired direction.

It is hence very important to understand how d-electron bands are affected by strain. In the past, strain has been proposed as a natural way to produce band engineering in oxides.\cite{qiao_gulec} In principle, in a cubic system strain has a detrimental effect by reducing orbital degeneracy and hence reducing the thermopower, particularly at high temperatures.\cite{prl_alex} To explore quantitatively the effects of strain and its related symmetry breaking in the actual values of the thermopower for typical oxides, we have chosen to model this using a very simple cubic perovskite oxide SrTiO$_3$, that is naturally n-doped. Those electrons will populate the t$_{2g}$ bands which will respond to strain in different ways depending on its magnitude, its sign, but more importantly, also depending on the doping level. We will analyze in detail the effects of strain not only in transport properties but also in terms of disturbing the electronic degeneracies in a t$_{2g}$ electron system and how these affect the thermopower and the effective mass of the system.

In principle, the findings obtained for SrTiO$_3$ should provide a general platform for understanding similar mechanisms that occur in other perovskite-related oxides of interest for TE applications.\cite{pablo_dp_arxiv}

\section{Computational procedures}


Ab initio electronic structure calculations based on the density functional theory (DFT)\cite{HK,KS} have been performed
using an all-electron full potential code ({\sc wien2k}\cite{WIEN2k}) on SrTiO$_3$.

The exchange-correlation term is parametrized depending
on the case. We have used the generalized gradient
approximation
(GGA) in the Wu-Cohen\cite{Wu_cohen_gga} scheme for structural optimizations (volume optimizations). For the electronic structure calculations we have used the semilocal potential developed by Tran and Blaha as a modification of the Becke-Johnson potential.\cite{TB-mBJ}  
This method typically allows the calculation of band gaps with an accuracy similar to the much more expensive GW or hybrid methods,\cite{TB-mBJ1} and in particular for 
SrTiO$_3$ it gives an accurate band gap.\cite{prl_alex} However, effective masses are not improved with respect to standard LDA/GGA functionals.\cite{mbj_meff}

The calculations were performed with a converged k-mesh and a value of  R$_{mt}$K$_{max}$= 7.0.
Spin-orbit coupling (SOC) was introduced in a second variational manner using the scalar
relativistic approximation.\cite{singhLAPW}
The $R_{mt}$ values used were in a.u.: 2.50 for Sr, 1.86 for Ti and 1.69 for O when studying SrTiO$_3$.

Transport properties have been calculated using the BoltzTrap\cite{boltztrap} code. This solves Boltzmann transport equation from first-principles calculations within the constant scattering time approximation. This strategy will allow us to obtain the Seebeck coefficient 
at different temperatures, which is, within that approximation, independent of the scattering time. These calculations require an even finer k-mesh for the Brillouin zone integrations ($25\times25\times25$).

We have simulated the effects of both tensile and compressive strains on SrTiO$_3$ by fixing the $a$ lattice parameter to that of several well-known systems typically used as substrates for 
thin film deposition (LAO: $a$= 3.821 \AA, LSAT: $a$=3.868 \AA, STO: $a$= 3.905 \AA, DSO: $a$= 3.942 \AA, GSO: $a$= 3.968 \AA)\cite{prl_alex,LAO_parameter,GSO_DSO_parameter} and relaxing the $c$ lattice 
parameter. 
We have thus explored how the electronic band structure evolves under different degrees of strain.

\section{Thermopower modeling}

In this section we will provide an expression for the thermopower as a function of temperature and strain and analyze the different temperature regimes. We will obtain the Seebeck coefficient for different strain situations from DFT calculations and interpret the results we get in terms of a simple model in order to make predictions easier. 
In the model that we propose, we have simplified the expression proposed by Kubo for the thermopower.\cite{kubo_formalism} Within this formalism, the equation to solve in order to obtain the Seebeck coefficient 
is the following:

\begin{equation}\label{kubo_thermopower}
 S=\frac{S^{(2)}/S^{(1)}+\mu /e}{T}
\end{equation}

where $\mu$ is the chemical potential, $e$ the absolute value of the electron charge, $T$ the temperature and $S^{(2)}$ and $S^{(1)}$ are integrals of time correlation functions. We will not go further 
in the discussion of these integrals since their shape makes them impractical when calculating transport properties in most cases. For this reason, 
in our model, we will replace these integrals by a simpler function of temperature.

Equation (\ref{kubo_thermopower}) can be split in two parts. 
The term $S^{(2)}/TS^{(1)}$ is the energy-transport term, while $\mu /Te$ is the entropy term, that will dominate at high temperature.\cite{chaikin1976thermopower} We can assume that the thermopower will follow the simple expression:

\begin{equation}\label{model_thermopower}
 S=\frac{A}{T+T_0}+S_{\left( T\rightarrow \infty \right)}
\end{equation}

where the first term on the right is dominant at low temperatures and corresponds to the energy-transport term, whereas the second term, that is associated to the entropy term, dominates at high enough temperature.

In the following subsections we will discuss in more detail the meaning of each term separately and the parameters that appear in our model.

\subsubsection{Low temperature Seebeck coefficient}



To interpret the origin and shape of the energy-transport term, we will make use of the results obtained from our ab initio calculations solving the Boltzmann transport equations to obtain the values of the constants $A$ and $T_0$ and try to provide a physical interpretation for them. This will be fit to the following expression:

\begin{equation}\label{fit_boltztrap}
 S=\frac{A}{T+T_0^{\prime}}+B
\end{equation}

in order to obtain the value of the parameter $A$, that accounts for the temperature-dependence of the energy transport term. 


The $T_0$ parameter is introduced in our model to make the Seebeck coefficient to be zero when temperature is zero. From eq. (\ref{model_thermopower}) it is easy to see that:

\begin{equation}\label{T_0_value}
 T_0=\frac{-A}{S_{\left( T\rightarrow \infty \right)}}
\end{equation}


Let us now try to analyze the physical meaning of the $A$ parameter relating it to the effective mass $m^*$. Our aim is not to give an exact value for the effective mass (since this would be 
a band-dependent and largely anisotropic in momentum space tensor\cite{janotti_vdwalle_sto}) from $A$, but just analyzing its variation with strain.

On the one hand, thermopower's Mott formula for semiconductors\cite{Fritzsche_semi_therm} has a similar shape to that of the model we are considering:

\begin{equation}\label{mott_formula}
 S_{Mott}=\frac{E_F-E_C}{eT}-\frac{k_B }{e}a_c
\end{equation}

where $E_F$ is the Fermi level, $E_C$ the minimum in energy of the conduction band (it will be at the $\Gamma$ point in STO), $k_B$ the Boltzmann constant and $a_c$ is a 
constant that depends on the material. Comparing with our model:

\begin{equation}\label{equivalence_mott_A}
 A=\frac{E_F-E_C}{e}
\end{equation}

On the other hand, if we consider an energy band with a parabolic dispersion:

\begin{equation}\label{parabolic_dispersion}
 E(k)=E_C+\frac{\hbar k^2}{2m^*}
\end{equation}

it can be shown that the Fermi level is given by:\cite{kittel_solids}

\begin{equation}\label{free_electron_gas}
 E_F=E_c+\left( \frac{3n}{\pi}\right)^{2/3}\left( \frac{h}{8m^*}\right)
\end{equation}

where $n$ is the carrier concentration and $h$ the Planck constant. Equation (\ref{free_electron_gas}) is the usual expression of the Fermi energy as a function of the carrier 
concentration for electrons in parabolic bands.

We can now consider together eq. (\ref{equivalence_mott_A}) and eq. (\ref{free_electron_gas}). Straightforward algebra allows to obtain an expression for the effective mass 
$m^*$ in terms of the parameter $A$, that we can obtain from a fit to our ab initio calculations of the thermopower:

\begin{equation}\label{effective_mass}
 m^*=\left[\frac{\left( \frac{3n}{\pi}\right)^{2/3}\left( \frac{h}{8m_0}\right)}{Ae} \right]m_0
\end{equation}

where $m_0$ is the bare electron mass. As we have mentioned previously, this effective mass corresponds to a simplified parabolic band picture, but it would be helpful to analyze its dependence on strain.


\subsubsection{High temperature Seebeck coefficient}

In this part, we will explain where the high temperature Seebeck coefficient comes from.
In Ref. \onlinecite{chaikin1976thermopower} an equation that relates the thermopower $S$ at high temperatures to the number of possible configurations $g$ is deduced:

\begin{equation}\label{seebhighT}
 S_{\left( T\rightarrow \infty \right)}=-\frac{k_B}{e}\frac{\partial \log g}{\partial N} 
\end{equation}

where $k_B$ is the Boltzmann constant, $e$ is the electron charge and $N$ is the number of electrons. In this paper the authors obtain the Seebeck coefficient 
in different situations. We will show some of the results in order to clarify the equation that we will use.

For a system of $N$ spinless fermions with $N_A$ sites to be occupied, 
the number of possible configurations $g$ is:

\begin{equation}\label{spinless}
 g=\frac{N_A!}{N!(N_A-N)!}
\end{equation}

using equation (\ref{seebhighT}), the so-called Heikes formula is obtained:

\begin{equation}\label{heikesfor}
 S_{\left( T\rightarrow \infty \right)}=-\frac{k_B}{e}\log \left[ \frac{(1-\eta)}{\eta} \right]
\end{equation}

where $\eta=N/N_A$. We want now to find an equation similar to (\ref{heikesfor}), but for the case of fermions with spin. We will be interested in the case in which the repulsion energy between 
fermions (for our purpose these fermions will be electrons) is larger than the thermal energy, so in each site there can be just one particle. The number of possible configurations 
is the one seen for spinless fermions (\ref{spinless}) multiplied by a $2^N$ factor that takes into account the spin degeneracy. Hence, we get:

\begin{equation}\label{withspin}
 g=\frac{N_A!}{N!(N_A-N)!}2^N
\end{equation}

so the Seebeck coefficient obtained is:

\begin{equation}\label{seebspin}
 S_{\left( T\rightarrow \infty \right)}=-\frac{k_B}{e}\log \left[ 2\frac{(1-\eta)}{\eta} \right]
\end{equation}

\begin{figure}[ht!]
\begin{center}
\includegraphics[width=\columnwidth,draft=false]{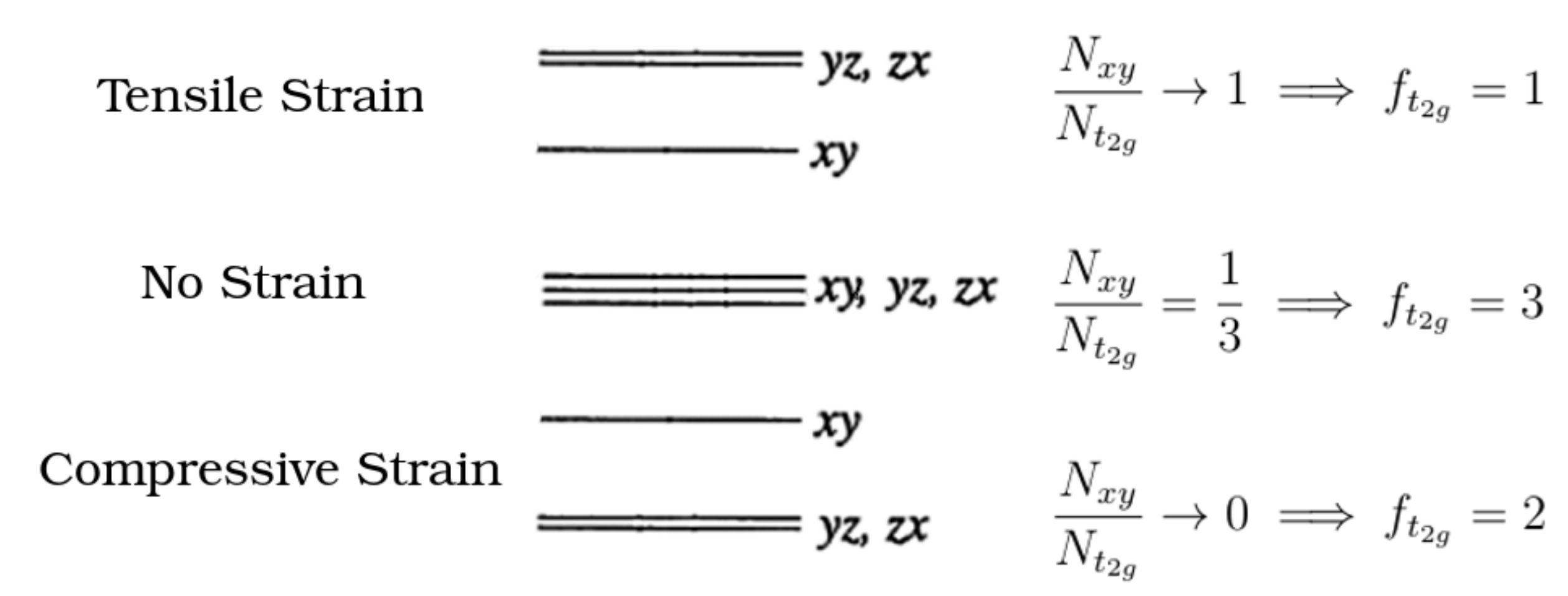}
\caption{We show the three limiting cases for the $t_{2g}$ degeneracy. In the case without strain (middle panel), all the orbitals are degenerate ($f_{t_{2g}}$= 3). If strain is applied, degeneracy is broken towards having a lower lying singlet 
(tensile strain: $f_{t_{2g}}$= 1, top panel ) or a lower lying doublet  (compressive strain: $f_{t_{2g}}$= 2, bottom panel ).} 
\label{lim_fact}
\end{center}
\end{figure}

Now that we have seen how to introduce the spin degeneracy, we are in the position to expand equation (\ref{withspin}) to account for the 
$t_{2g}$ orbital degeneracy that occurs at the bottom of the conduction band in electron-doped STO. We will introduce heuristically a factor $f_{t_{2g}}$ in eq. (\ref{withspin}), that weighs the $t_{2g}$ degeneracy in a similar way as the factor $2$ 
does for the spin degeneracy:

\begin{equation}\label{t2gdeg}
 g=\frac{N_A!}{N!(N_A-N)!}2^Nf_{t_{2g}}^N
\end{equation}

so inserting this in equation (\ref{seebhighT}) we obtain:

\begin{equation}\label{seebt2g}
 S_{\left( T\rightarrow \infty \right)}=-\frac{k_B}{e}\log \left[ 2f_{t_{2g}}\frac{(1-\eta)}{\eta} \right]
\end{equation}

but we have not talked about the shape of this factor $f_{t_{2g}}$ yet. Our goal is to find an expression that reproduces the degeneracy behavior 
in different strain situations. We will model it as a function of the number of electrons in the $d_{xy}$ orbital, $N_{xy}$ and the total number of electrons in the 
$t_{2g}$ manifold, $N_{t_{2g}}$. Figure \ref{lim_fact} shows the three limiting cases we will have to consider in order to model the degeneracy in every case. The unstrained case implies all t$_{2g}$ bands are degenerate, and hence there is a triple orbital degeneracy. If tensile (compressive) strain occurs, the $d_{xy}$ ($xz/yz$) band/(s) lies (lie) lower in energy and then a single (double) degeneracy occurs.

So we encounter that the factor $f_{t_{2g}}$ must be a function of $x=N_{xy}/N_{t_{2g}}$, its value must be contained in the interval $[1,3]$, and the maximum being equal to $3$ at $x=1/3$. Considering all these facts, we 
are going to fit the points shown in Fig. \ref{lim_fact} to the following function (we chose a Gaussian because it is the simplest smooth function that one can think of that has only one adjustable extremum in the $[1,3]$ interval):

\begin{equation}\label{ffactor}
 f_{t_{2g}}=\gamma+\frac{\alpha}{\beta\sqrt{\frac{\pi}{2}}}e^{-2\left(\frac{(x-1/3)}{\beta}\right)^2}
\end{equation}

where the constants $\alpha$, $\beta$ and $\gamma$ are completely determined by the three limiting cases shown in Fig. \ref{lim_fact}.

\begin{figure}[ht!]
\begin{center}
\includegraphics[width=\columnwidth,draft=false]{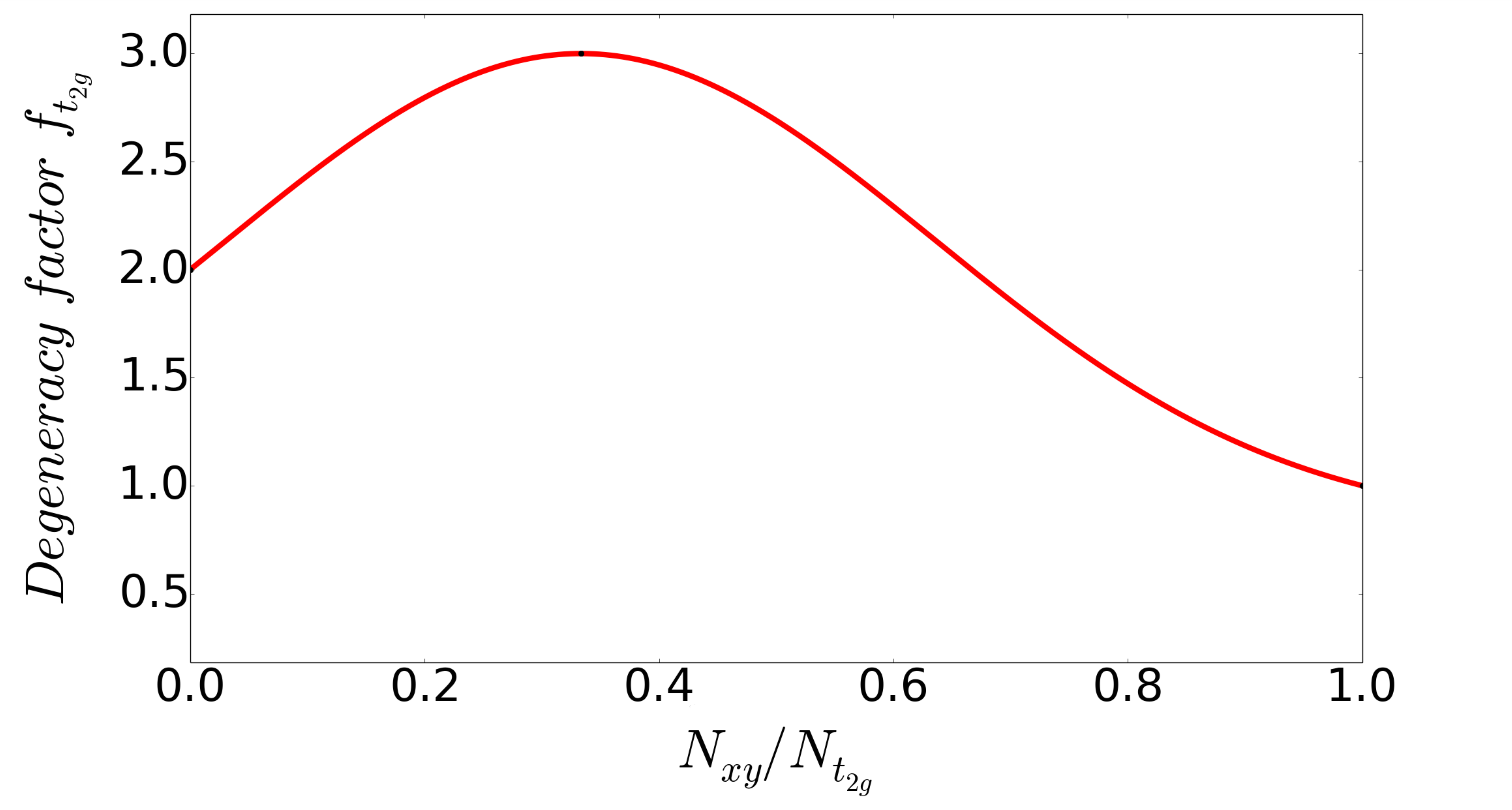}
\caption{(Color online.)$f_{t_{2g}}$ function that results from eq. \ref{ffactor}. This degeneracy factor fulfills the three limiting cases analyzed in Fig. \ref{lim_fact} for the effect of strain in a t$_{2g}$ electron system. It accounts for the orbital degeneracy of the system as a function of strain. N$_{xy}$ and N$_{t_{2g}}$ need to be determined ab initio.} 
\label{strain_deg}
\end{center}
\end{figure}

A plot of the factor as a function of $x$ can be seen in Fig. \ref{strain_deg} .

We have found an analytical function for the high temperature thermopower from a statistical calculation. A degeneracy factor for the $t_{2g}$ was introduced as a function of $N_{xy}/N_{t_{2g}}$. 
This procedure will make possible to calculate the entropy-related term $S_{\left( T\rightarrow \infty \right)}$ from our DFT calculations just by computing $N_{xy}/N_{t_{2g}}$ for each strain introduced in the STO structure.

What we have done so far is being able to decouple the influence in thermopower variations with strain that come from electronic degeneracies and from band structure modifications such as effective mass changes, that strain will also introduce.

\section{S\MakeLowercase{r}T\MakeLowercase{i}O$_3$ calculations}

We have run calculations in SrTiO$_3$ simulating different strain situations by fixing the lattice parameter $a$ to that of the different standard perovskite substrates mentioned in 
the computational procedures section. For each $a$, we have optimized the lattice parameter $c$. We have not considered oxygen octahedral tilts, since this kind of distortions are 
hardly dependent on the impurity introduced as dopant. Therefore, we have limited our analysis to a cell volume distortion caused by biaxial strain. We have found that tensile (compressive) 
strain increases (reduces) the unit cell volume (see Fig. \ref{bwcellvol} in Appendix \ref{AppA}). This has been experimentally reported previously for other oxides.\cite{petrie_strain_2016}

With these structures we have performed electronic structure calculations including spin-orbit coupling with the TB-mBJ exchange-correlation potential, 
that allows to give an accurate band gap without additional computational cost. Band structure and density of states plots are shown in Fig. \ref{bandDOS} in Appendix \ref{AppA} for three different strain cases.
We have found that the $t_{2g}$ bandwidth reduces with the unit cell volume and consequently with tensile strain (see Fig. \ref{bwcellvol} in Appendix \ref{AppA}). This would have an implication on the effective mass because a reduced bandwidth leads to an increased average effective mass.

We have calculated the number of electrons in the $d_{xy}$ orbital, $N_{xy}$ and the total number of 
electrons in the $t_{2g}$ manifold, $N_{t_{2g}}$. Both values were obtained by integrating inside the Ti muffin-tin sphere. We will assume that the degree of localization will be identical for the three t$_{2g}$ orbitals and hence integrating inside the muffin-tin spheres is sufficient for our purposes since we are only interested in the ratio. This ratio is needed to obtain the degeneracy factor $f_{t_{2g}}$ and consequently the thermopower at high temperature $S_{\left( T\rightarrow \infty \right)}$ 
as a function of doping.

\begin{figure}[!ht]
\begin{center}
\includegraphics[width=\columnwidth,draft=false]{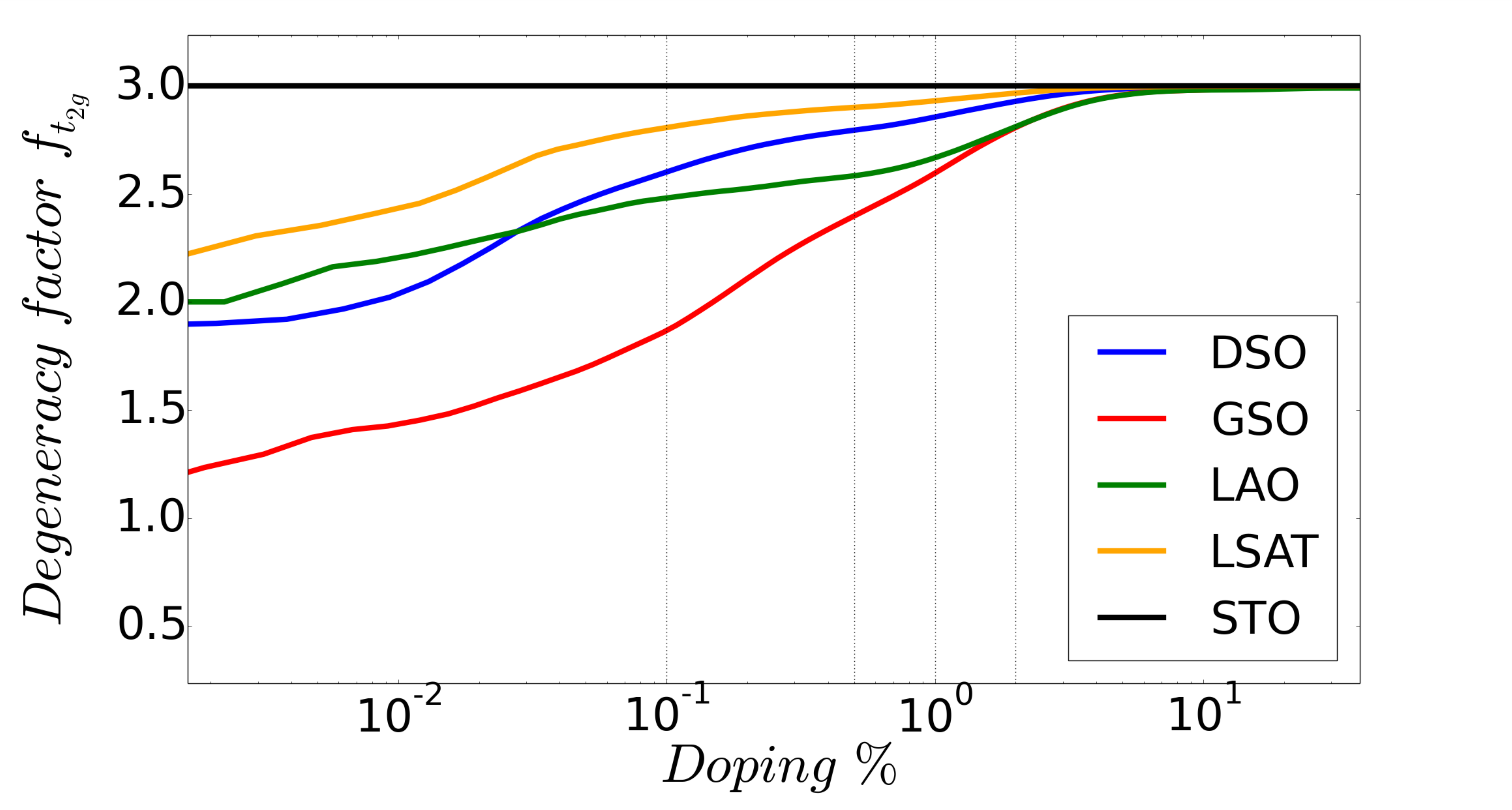}
\end{center}
\caption{(Color online.) Degeneracy factor as a function of doping. In the low doping regime the degeneracy factors collapse to the three limiting cases exposed in Fig. \ref{lim_fact}, so they become largely dependent on strain. 
As we increase the doping level the factor becomes strain independent and tends to $3$ as the t$_{2g}$ bands become more populated. Our calculations show that this occurs between 2-5 \% Nb-doping.  
In the case of unstrained STO we see that the degeneracy factor is completely independent of the doping level since there is always a triple degeneracy in that case. This figure suggests that degeneracy effects vanish above about 5\% doping.} 
\label{sto_deg}
\end{figure}

\begin{figure*}[!ht]
\begin{center}
\includegraphics[width=\columnwidth,draft=false]{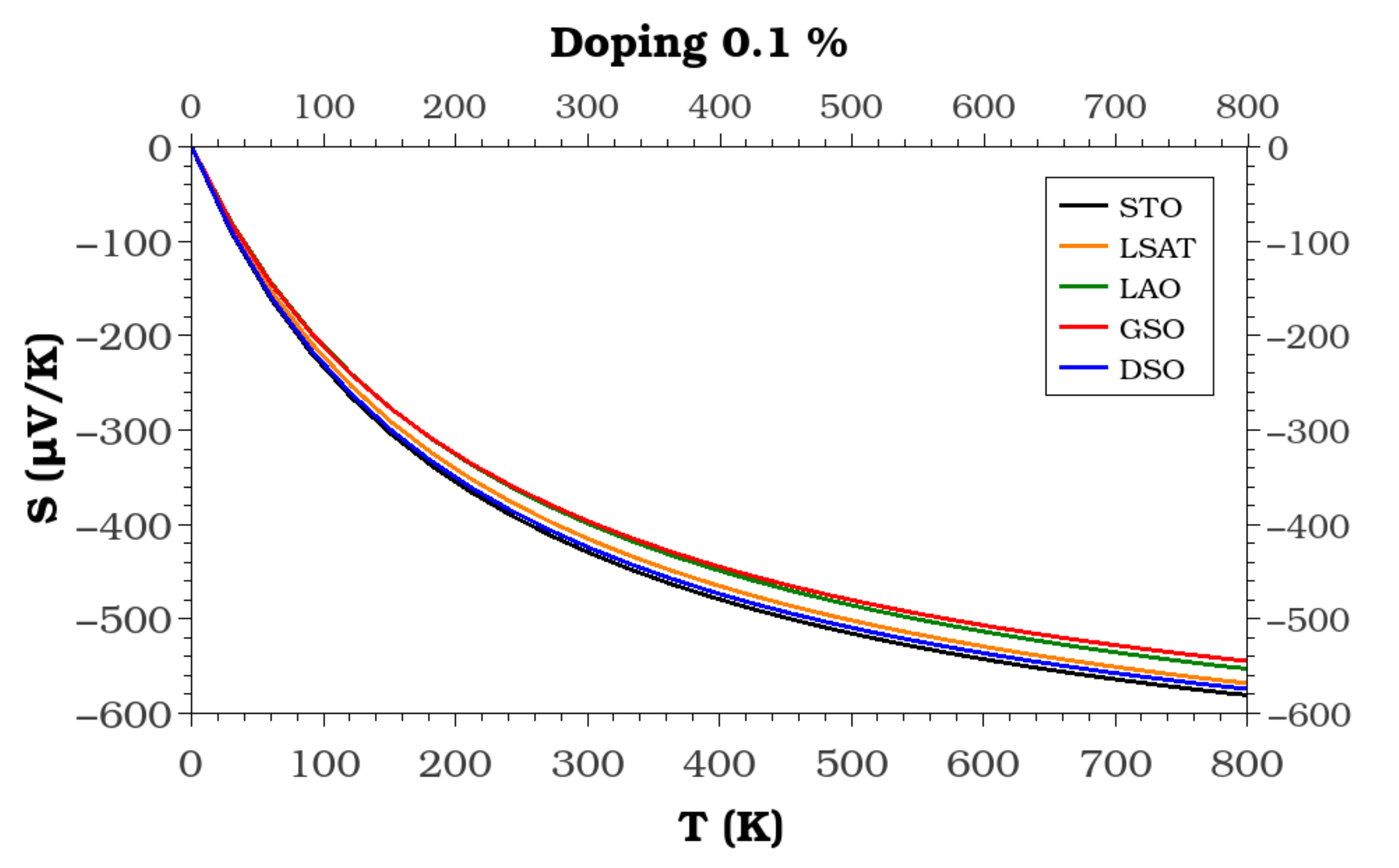}
\includegraphics[width=\columnwidth,draft=false]{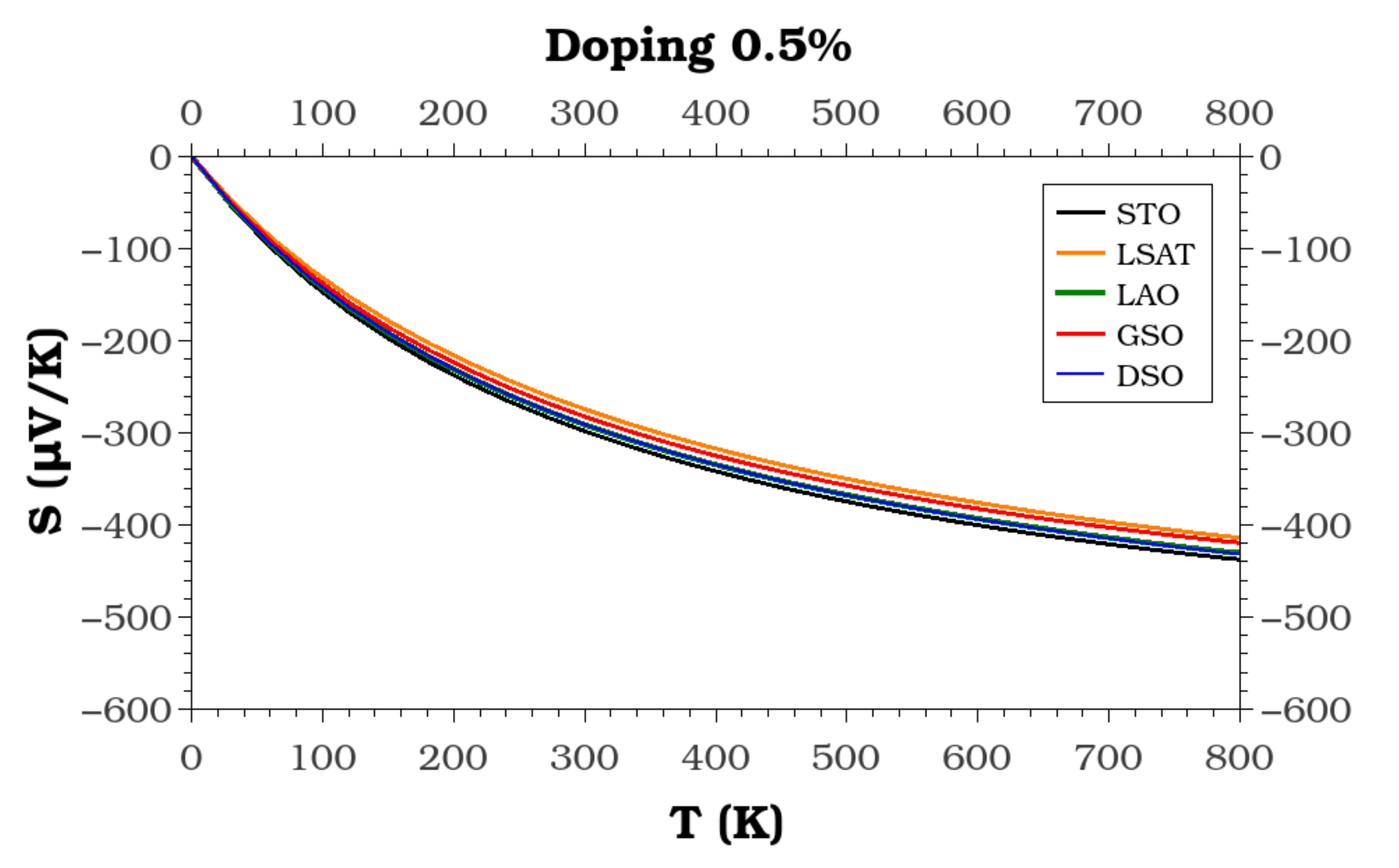}
\includegraphics[width=\columnwidth,draft=false]{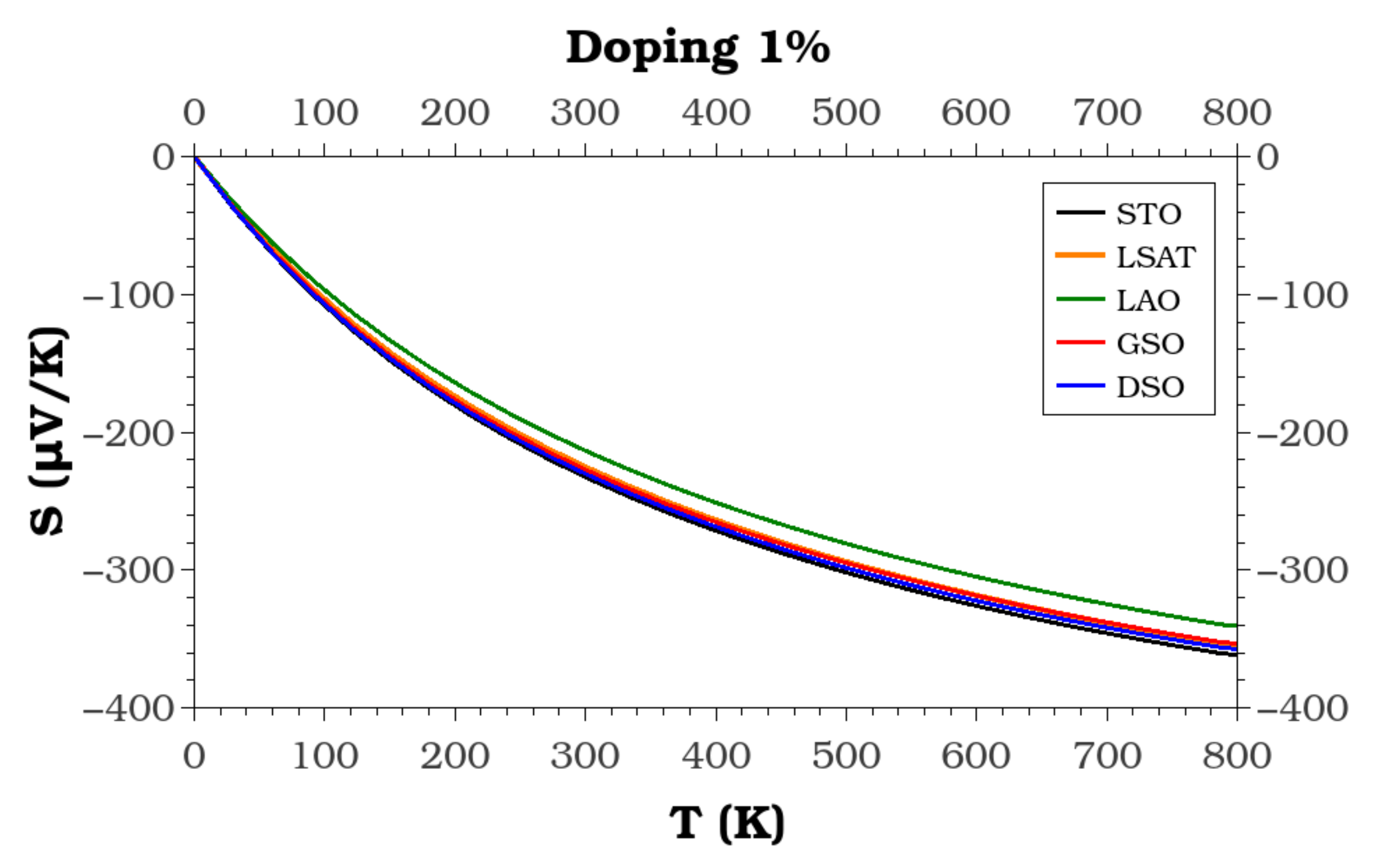}
\includegraphics[width=\columnwidth,draft=false]{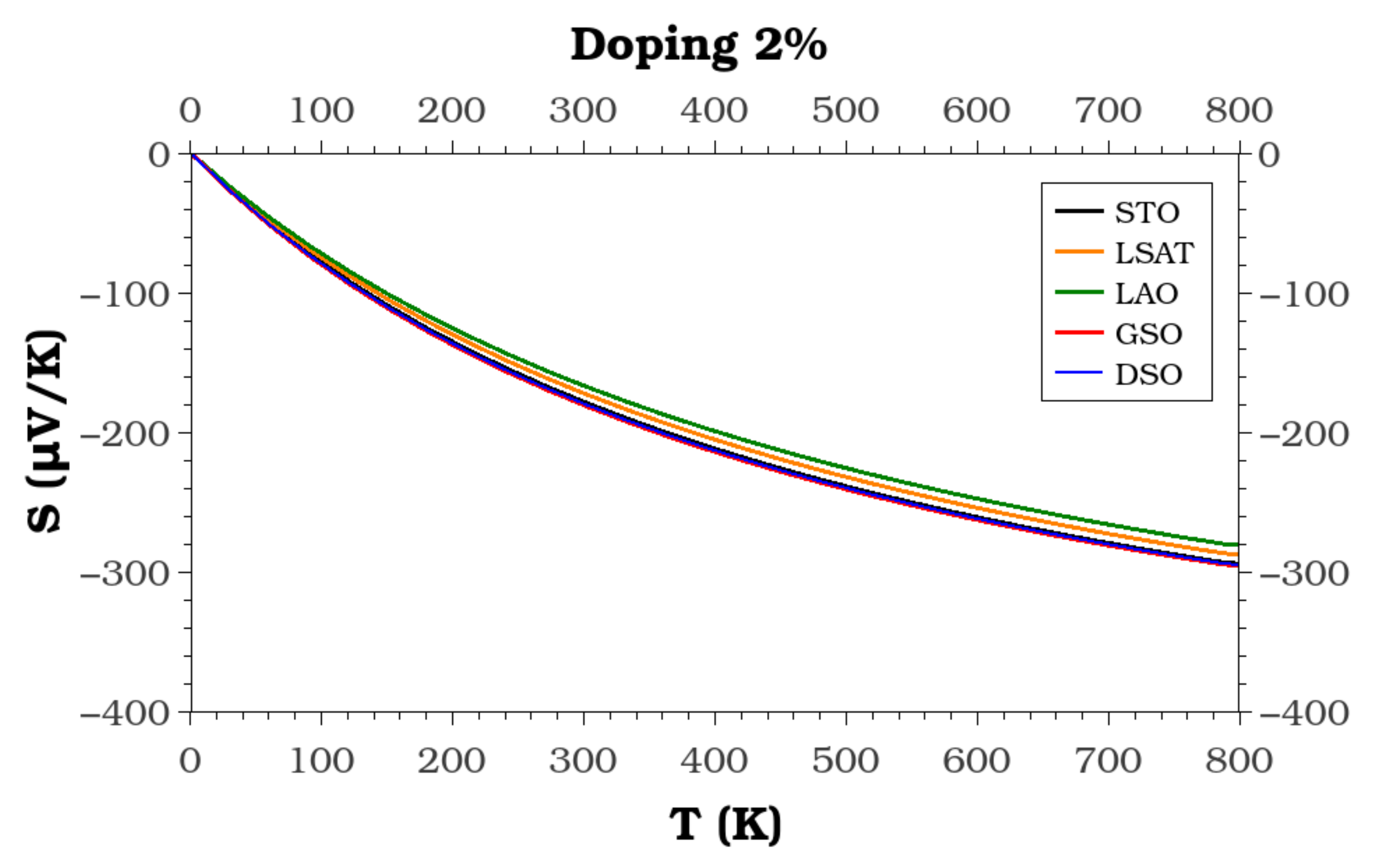}
\includegraphics[width=\columnwidth,draft=false]{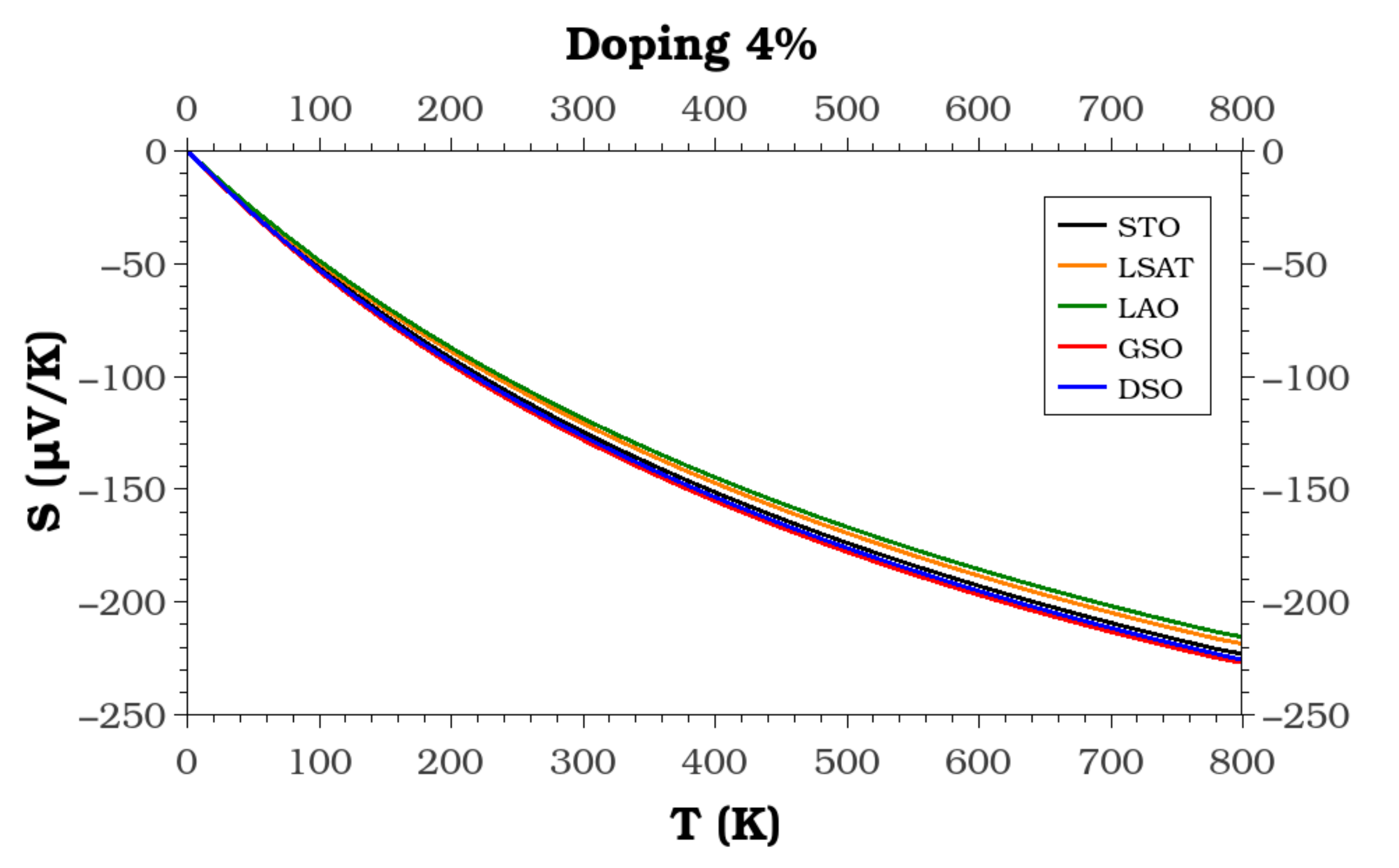}
\includegraphics[width=\columnwidth,draft=false]{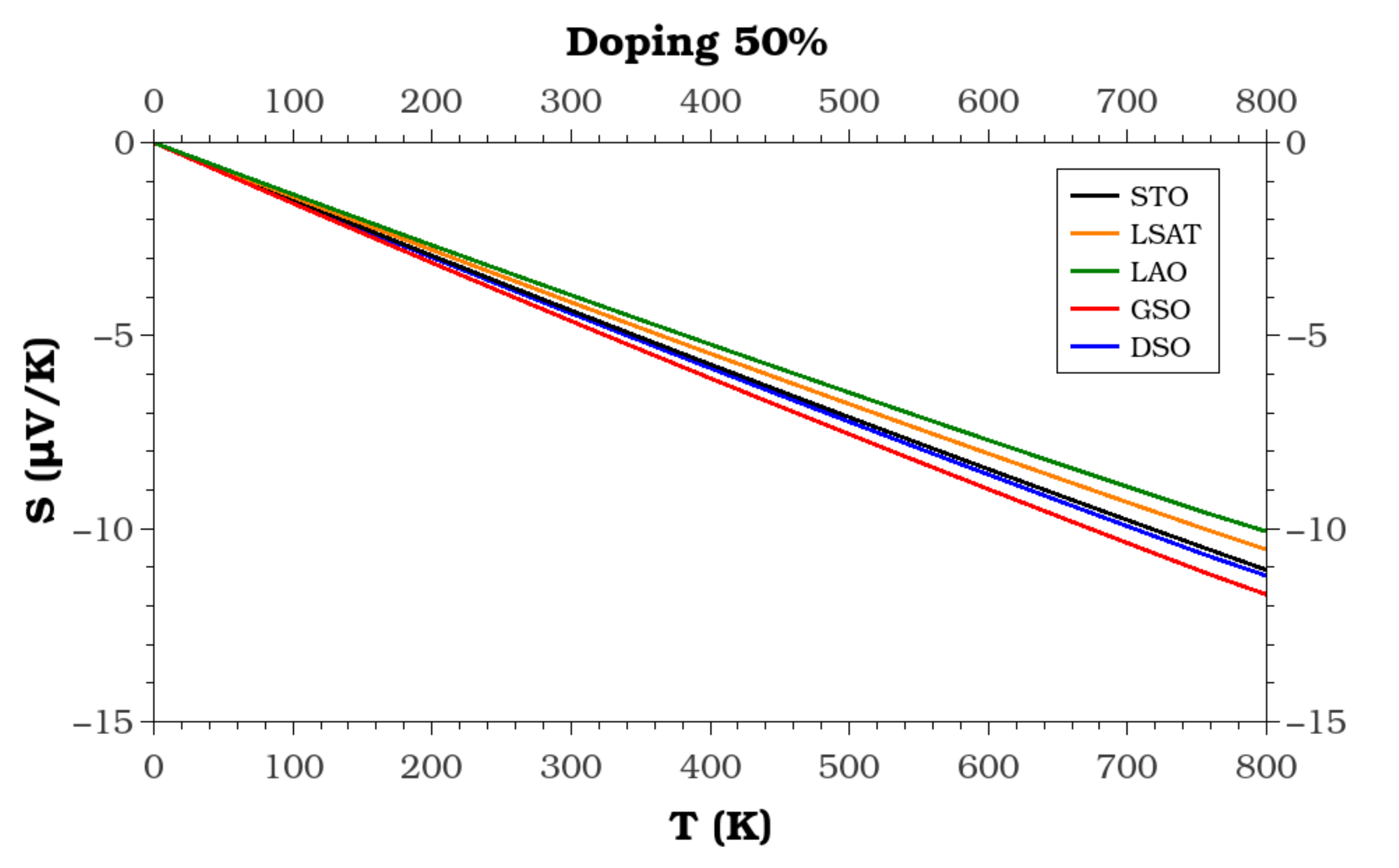}
\caption{(Color online.) Representation of the Seebeck coefficient as a function of temperature using eq. (\ref{model_thermopower}) for six different doping regimes. 
The parameters obtained in each case are compiled in Table \ref{parameters} in Appendix \ref{AppA}. It can be seen that in the low doping regime (below 2.0\%) the entropy-related term dominates the evolution with strain, leading to a reduction in the Seebeck coefficient (in magnitude) caused by any kind of strain. However, for high doping, it is the 
low temperature (energy-transport term) thermopower the one that dominates the strain tendency because $S_{\left( T\rightarrow \infty \right)}$ becomes strain independent (see Fig. \ref{sto_deg}). In that case, we see that tensile strain increases the thermopower (in magnitude).}
\label{seebeck_evo}
\end{center}
\end{figure*}

Figure \ref{sto_deg} shows the degeneracy factor dependence with doping for the different strain situations analyzed. We can see there, as a function of electron doping 
(in terms of electrons doped per unit cell, or Nb atoms substituting Ti, thus adding one extra electron to the conduction band per dopant). We can see that at low doping, the three 
limiting cases we discussed above are obtained (in the pure single ion picture), with the unstrained case having threefold degeneracy and the compressive or tensile strain cases 
tending to a twofold degenerate state or a singlet, respectively. As doping increases, the t$_{2g}$ bands become more heavily populated and the actual band splittings become less and 
less important. Above 5\% doping, the degeneracy factor becomes very close to 3, almost independent of strain. This illustrates that, even though in principle strain can tune 
degeneracies (by removing them), thermopower at high temperatures might not be affected if doping is substantial. This limit will of course depend on the system, but for 
STO we see that it is obtained (within a rigid band approximation) at values which are on the order of those sought for in the case of TE applications ($\sim$ 10$^{20}$ cm$^{-3}$). The important point to notice here is that the effect of strain on 
STO will not damage the TE efficiency at high temperature by reducing orbital degeneracies, when working at those high doping levels we are discussing.

Introducing these results of the degeneracy factor $f_{t_{2g}}$ in eq. (\ref{seebt2g}), we can obtain the thermopower at high temperature $S_{\left( T\rightarrow \infty \right)}$ as a function 
of doping for the five strain situations that we are analyzing. Once $S_{\left( T\rightarrow \infty \right)}$ is computed, we have to solve the Boltzmann transport equations to obtain the parameter $A$. After that, we are in the position to get the value of $T_0$ using eq. (\ref{T_0_value}) and consequently determine the three 
parameters of our model (eq. (\ref{model_thermopower})). We must warn the reader that the Boltzmann transport equations used to compute the parameter $A$ would be less accurate at lower temperatures (in particular the phonon drag term is not included). Consequently, 
we will restrict our conclusions to analyze the evolution with strain and not focus too much on the actual values predicted for the thermopower. Also, discrepancies with experimental values could be due to the use of the constant relaxation time approximation which is used routinely but could have its limitations.\cite{filippetti_relaxation_time}

Figure \ref{seebeck_evo} shows six representations of the thermopower obtained using the model that we propose (eq. (\ref{model_thermopower})) at different doping levels for the five 
strain cases analyzed, utilizing the values of the parameters that we have just calculated. The parameters $A$, $S_{\left( T\rightarrow \infty \right)}$, $T_0$ and also the obtained effective masses are summarized in Table \ref{parameters} of Appendix \ref{AppA}. 

\begin{figure}[!ht]
\begin{center}
\includegraphics[width=\columnwidth,draft=false]{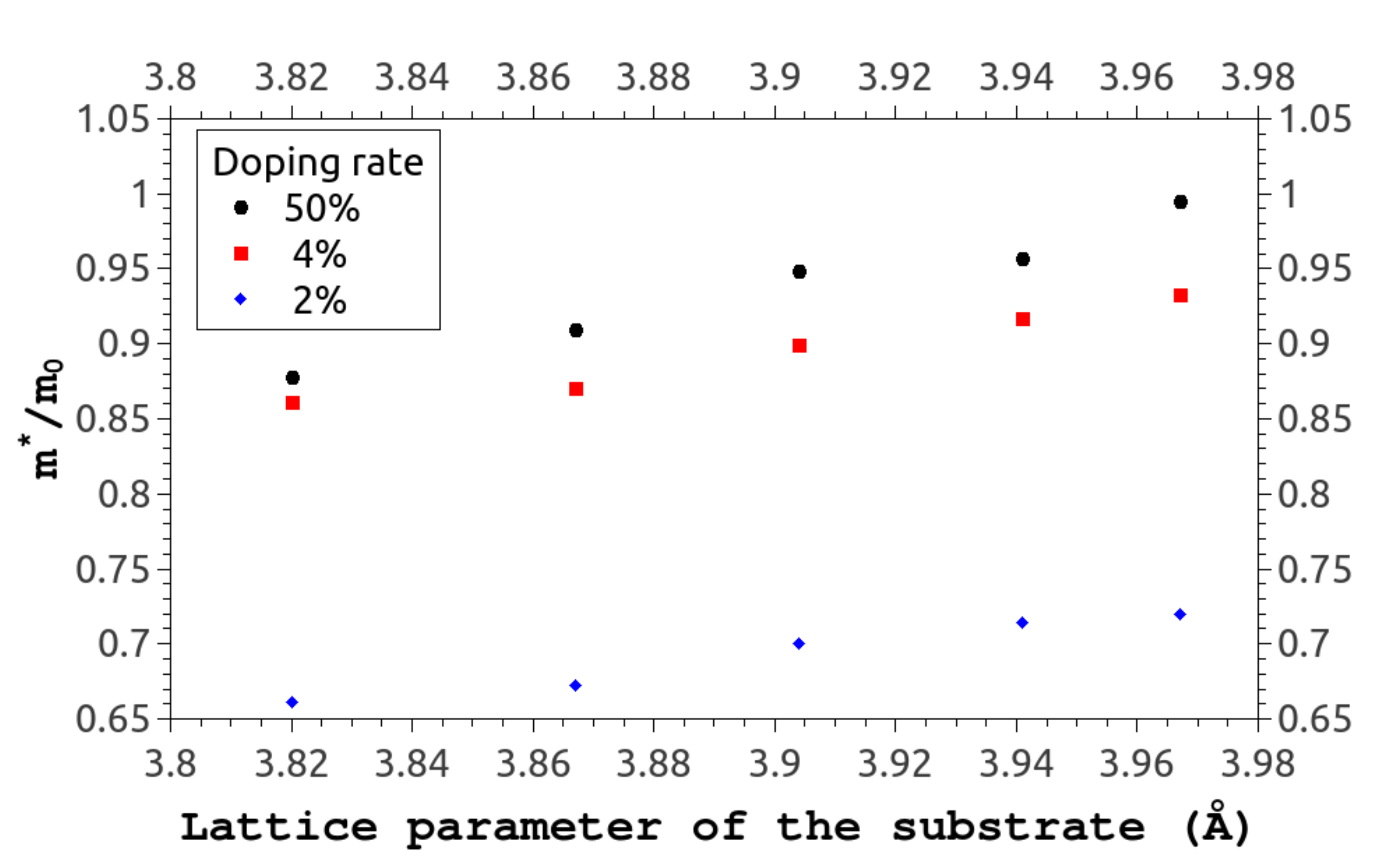}
\end{center}
\caption{(Color online.) Effective mass as a function of strain in the high doping regime. Effective mass increases (decreases) with tensile (compressive) strain and barely changes once all the bands become populated. The limit of 50\% doping, even though beyond the validity of the rigid band approximation is shown as the limiting case found in LaAlO$_3$/STO interfaces.} 
\label{effm_dop_strain}
\end{figure}

Figure \ref{seebeck_evo} shows that the high temperature thermopower decreases in absolute value as we increase the doping level, as one would expect. Figure \ref{effm_dop_strain} shows the dependence that effective masses have with strain for the high doping regime. We can observe there that the theoretical 
effective masses obtained increase their value as doping increases. However, eq. (\ref{effective_mass}) proposed to obtain the effective masses from parameter A is just a mechanism to get the effective mass tendency with strain. As we have said, it will only be valid in the high doping regime 
so we will focus on analyzing the evolution of the effective mass with strain but not so much on its actual doping dependence, since this will be quite dependent on the type of dopant. This feature is analyzed in Ref. \onlinecite{wunderlich_enhanced_2009} concluding 
that the effective mass will also depend on the kind of defect introduced to dope STO and the kind of distortions these introduce in the lattice.

Analyzing now each doping regime, we see that for low doping levels the strain tendency is given by the high temperature seebeck $S_{\left( T\rightarrow \infty \right)}$, i.e. we see in Fig. \ref{seebeck_evo}, for 
 $0.1\%$ doping ($\sim 1.7 \times 10^{19}$ cm$^{-3}$)that the Seebeck evolution with temperature follows the same strain dependence than $S_{\left( T\rightarrow \infty \right)}$. The unstrained case (with $a$ equal to that of STO) is the one that has the largest Seebeck value (in absolute value) at any temperature. 
Meanwhile, at high doping levels, greater than or equal to $2\%$, according to the results shown in Fig. \ref{sto_deg}, $S_{\left( T\rightarrow \infty \right)}$ stops being strain dependent. Consequently, the only dependence on strain in the evolution of the Seebeck with temperature comes from the $A$ parameter. 
We can easily check that the curves in the $2\%$, $4\%$ and $50\%$ doping levels ($\sim 3.3\times 10^{20}$, $\sim 6.6\times 10^{20}$ and $\sim 8.4\times 10^{21}$ cm$^{-3}$ respectively) follow the strain dependence given by parameter $A$ in Table \ref{parameters}, i.e., the inverse behaviour that effective masses have with strain (see Fig. \ref{effm_dop_strain}). This comes about due to the fact that f$_{t_{2g}}$ becomes very close to 3 and strain independent at those doping levels. 
Hence, the only changes to the thermopower occur via variations of the energy-transport term, in which (for high doping regimes $2\%$ or more) the thermopower is enhanced (in absolute value) by tensile strain.

For analyzing this dependence in detail, we have obtained the effective masses and its strain and doping dependence. Despite the fact that we show the effective masses calculated in the six cases (see table \ref{parameters}), it is hard to analyze its strain dependence away from the high doping limit. At $2\%$ doping, we can observe that the effective mass is larger for the unit cells with a larger volume (volume increases for tensile strain). Increasing the volume leads to a reduction in the hopping parameter, smaller band widths and hence a larger effective mass.\cite{ohta_large_2005} Above $2\%$ doping, such dependence with strain of the effective mass is retained. For lower doping values, the dependence with strain is more complex due to the partial involvement of different bands. The effective mass increases monotonously with doping and it reaches about 1.0 m$_0$ at about $50\%$ doping. In that doping region, a single  parabolic band picture makes more sense than at low doping, where the sole concept of a single effective mass where all the bands contribute in a similar fashion is more difficult to identify.


In Ref. \onlinecite{janotti_vdwalle_sto} Janotti et al. analyze the directional effective masses dependency with strain. In order to compare with their data, we can obtain a mean effective mass $m_{ave}^*$ performing a harmonic mean 
of the directional effective masses $m_{\Gamma -X}$ (parallel) and $m_{\Gamma -\overline{X}}$ (perpendicular to the strain direction) given in this article and we can compare these $m_{ave}^*$ with the 
effective masses we have obtained for $2.0\%$ doping, e.g. In our calculations LSAT corresponds to biaxial strain $-1.0\%$, STO $0.0\%$ and DSO $+1.0\%$. The results are summarized in Table \ref{eff_m_comp}. Their results, when analyzed this way, agree with ours in the strain dependence of the effective mass. However, the actual values are largely doping dependent, as we have discussed here.

\begin{table}[h!]
  \centering 
  \begin{tabular}{|c|c|c|c|c|}
 \hline
 Strain & \multicolumn{1}{l|}{$m_{\Gamma -X}$} & \multicolumn{1}{l|}{$m_{\Gamma -\overline{X}}$} & \multicolumn{1}{l|}{$m_{ave}^*$} & \multicolumn{1}{l|}{Our $m^*$ ($2\%$ doping)} \\ \hline
 $-1.0\%$ & 0.67 & 0.40 & 0.547 & 0.672 \\ \hline
 $0.0\%$ & 0.55 & 0.55 & 0.550 & 0.700 \\ \hline
 $+1.0\%$ & 0.42 & 2.2 & 0.575 & 0.714 \\ \hline
\end{tabular}
  \caption{Comparison between the effective masses obtained by A. Janotti et al. [\onlinecite{janotti_vdwalle_sto}] and the ones from our model at 2$\%$ doping. The masses are given in units of the free electron mass $m_0$. As we explain in the text, the actual values for the effective masses are largely doping dependent, but the evolution with strain we obtain is retained, and consistent with previous results in the literature.}
  \label{eff_m_comp}%
\end{table}%

The effective masses are slightly larger in our model than the ones provided in Ref. \onlinecite{janotti_vdwalle_sto}. We are comparing our calculations at $2.0\%$ doping level with the effective masses calculated from the inverse of the second 
derivative of the first conduction band in their case. However, the strain tendency is the same 
for both calculations. In view of the last two columns of Table \ref{eff_m_comp} we can conclude that the mean effective mass in bulk STO increases (decreases) with positive (negative) biaxial strain. 
The same dependence is obtained in Ref. \onlinecite{wunderlich_enhanced_2009} where they found that an increase of the effective mass produces an increase of the thermopower.

A  similar analysis to this one could be performed to check the Seebeck dependence with strain that we have obtained in our model. In Ref. \onlinecite{zou_effect_2013} Daifeng Zou et al. 
calculate the strain dependence of the thermopower in different 
directions (parallel and perpendicular to the strain plane). A harmonic mean value of their results provides the same strain dependence that we have obtained from our model, i.e., tensile strain enhances the thermopower. 

The importance of the dependence of the Seebeck coefficient with strain is crucial when fine-tuning the TE properties of oxides, which often comes about due to nanostructurization, where strain is a key factor. This simple model we have put forward, supported by ab initio calculations, allows to account separately for the different factors that will influence the thermopower. In principle, one could naively think that applying strain simply destroys the t$_{2g}$ degeneracy ruining the high temperature Seebeck coefficient, but we can see from our calculations on STO that this effect is small since the usual degrees of strain attained experimentally do not fully break the t$_{2g}$ degeneracy, in particular at substantial doping ($\geq 10^{20}$ cm$^{-3}$), as is required for TE applications.\cite{ohta_recent_2008} This is important in case one thinks of oxide-based design of new TE materials and wants to tune its thermopower with strain. We can see that the band effects such as band alignment modifications, doping levels and effective masses are much more important in affecting the Seebeck, and these need not act in a detrimental way for TE properties. Of course, the doping limit that leads to washing away the orbital degeneracies will be largely system-dependent.

\section{Concluding remarks}

In conclusion, we have analyzed the thermopower in STO and its strain dependence for different n-doping levels and have been able to decouple the different origins of these dependencies: those coming from electronic degeneracies and entropy-related thermopower to those coming from the actual modifications in the band structures.

In the first part of this work, we have proposed a simplified model for the Seebeck coefficient. We consider the Kubo formalism and its temperature limits to build a simple equation for 
the thermopower based on the Mott formula. The first part of this equation corresponds to the energy-transport term and dominates in the low temperature regime. We obtain also a relationship between the free parameter of this term and a theoretical 
effective mass. The second term in our model corresponds to the entropy term at high temperatures, which is given by a Heikes-like equation. We introduce a degeneracy factor in this equation to model 
the effect of strain on the high temperature thermopower by affecting orbital degeneracies in the $t_{2g}$ manifold. We show how this degeneracy factor can be modeled by the use of ab initio calculations. 
Thus, we provide a way to model the Seebeck coefficient in a simple way by understanding how the effective mass and the electronic degeneracies evolve with strain.
In the second part of the work, we have applied the model obtained in the first part to analyze the thermopower strain dependency of STO. We performed DFT-based calculations varying the lattice parameter $a$ matching it to that of five 
different usual substrates. The parameters that appear in our model were calculated for these five strain cases and for six different doping values. Thus, we obtained the thermopower as a function 
of temperature for different values of strain and doping. The results of our model can be summarized as follows: i) at low doping any type of strain will decrease the thermopower. 
This is caused by the dominance of the entropy-related term, since strain always reduces the degeneracy factor. 
ii) at high doping, the one useful for TE applications, starting at $2\%$ Nb-doping ($\sim 10^{20}$ cm$^{-3}$), the energy-transport term is dominant.  
Thus, at operating temperatures, tensile strain enhances the thermopower. In short, we have obtained that in the low doping regime the electron degeneracy dominates the Seebeck, while 
in the high doping regime the cell volume effect dominates (increasing volume leads to an increase in the Seebeck). In principle, this kind of reasoning can be extended to other 
oxides where orbital degeneracy plays a role. However, the particular results of a similar analysis such as the doping values of the various limits, will be largely system-dependent.

\acknowledgments

This work was supported by Xunta de Galicia under the Emerxentes Program via the project no. EM2013/037 and the MINECO via project MAT2013-44673-R. V.P. acknowledges support from the MINECO of Spain via the Ramon y Cajal program (RyC-2011-09024). We thank P. Villar Arribi for fruitful discussions. We have benefited greatly from discussions with E. Ferreiro-Vila, A. Sarantopoulos and F. Rivadulla.

\appendix

\begin{figure}[!hb]
\begin{center}
\caption*{\section{}\label{AppA}}
\includegraphics[width=\columnwidth]{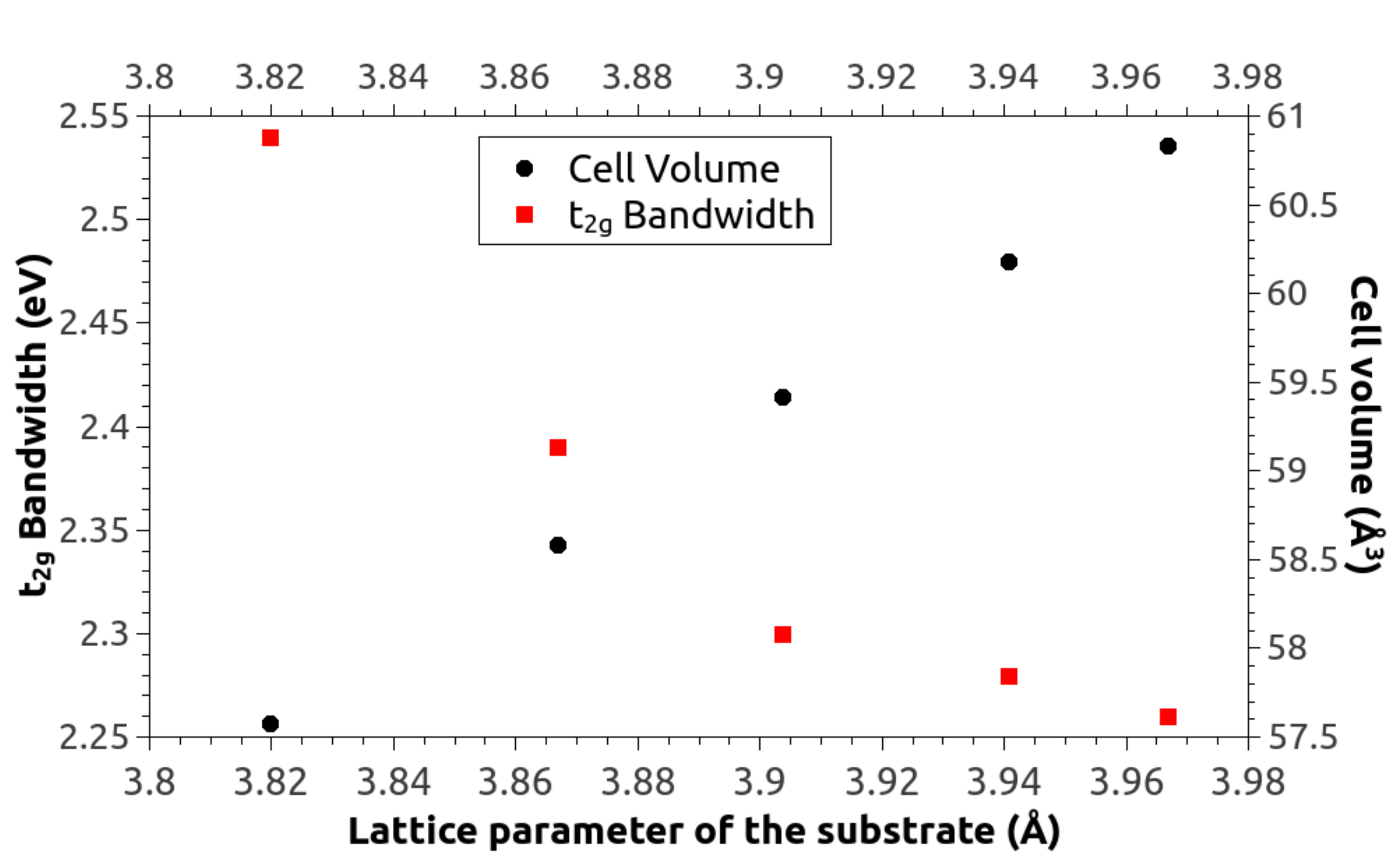}
\end{center}
\caption{(Color online.) Bandwidth and unit cell volume for different strain situations. We see that bandwidth decreases with volume and volume is increased as positive (tensile) biaxial strain is applied.} 
\label{bwcellvol}
\end{figure}

In this Section we provide a few extra details for completeness of the electronic structure analysis. In Fig. \ref{bwcellvol} we plot the evolution of the unit cell volume with strain. We observe that, as discussed throughout the text, tensile strain leads to an increase in unit cell volume. The volume reduction that occurs for compressive strain leads to a smaller Ti-Ti distance, which increases the hopping integrals and hence leads to an increased bandwidth. This bandwidth increase should be related to a decrease in effective mass, as has been explained throughout the main text. Our analysis does not include, as we have also cautioned the readers, the specificities of the different types of dopants STO can sustain, we are only considering the pure effect of strain without the additional local distortions introduced by point defects and/or impurities.

\begin{figure}[!ht]
\begin{center}
\includegraphics[width=\columnwidth]{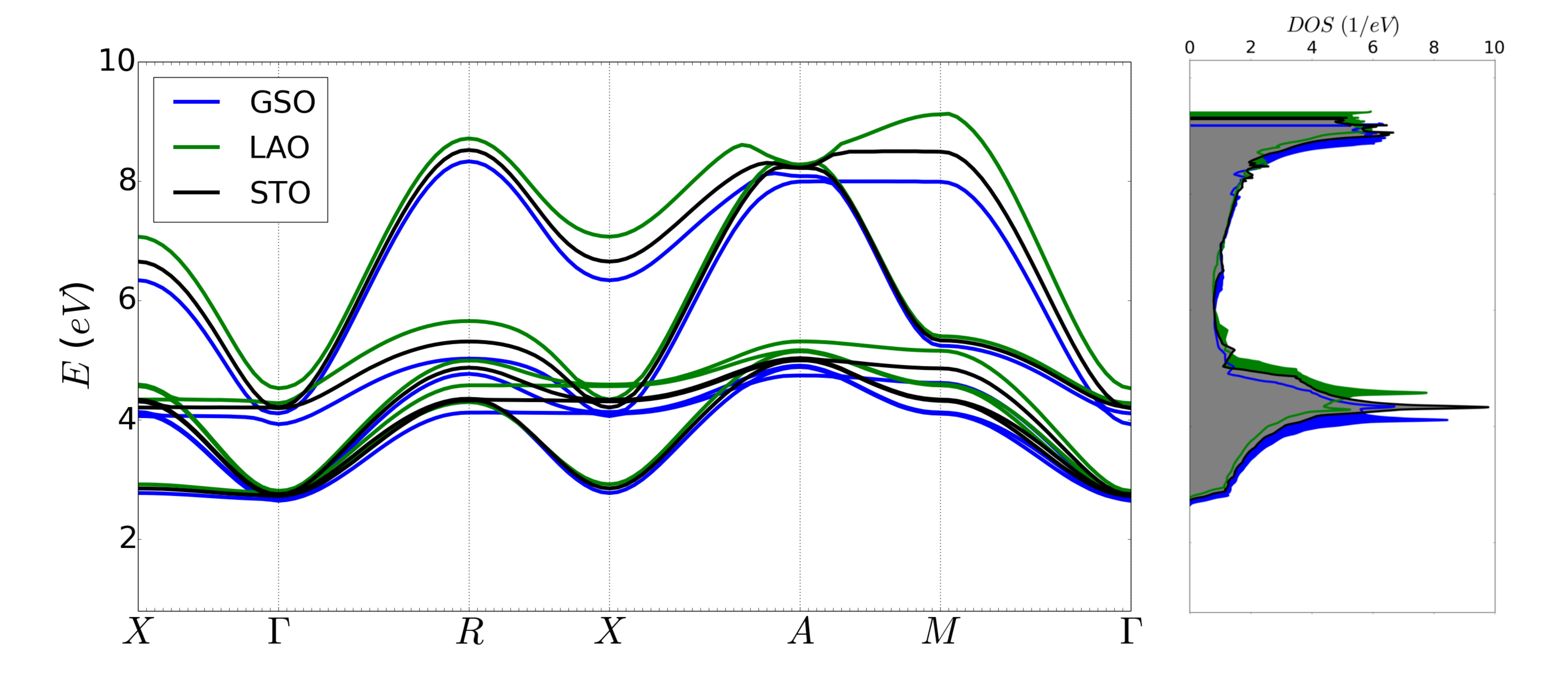}
\end{center}
\caption{(Color online.) Band structure representation and density of states (DOS) calculated for different strain values. It can be seen how the bandwidth increases for compressive strain and decreases for tensile strain.} 
\label{bandDOS}
\end{figure}

Figure \ref{bandDOS} shows the band structures (only the conduction bands close to its bottom) for different strain situations together with the corresponding DOS's on the same energy scale. GSO corresponds to the tensile strain limit and LAO to the compressive one. We can observe that the former leads, as explained above, to a reduction of the t$_{2g}$ bandwidth, together with some shifts in the DOS peaks. The bottom of the conduction band is at $\Gamma$. We can see also that the position of the unoccupied t$_{2g}$ bands is higher in energy for compressive strain and lower for the tensile strain case.

\begin{table*}[!ht]
  \centering 
  \caption{Model's parameters for the six representative doping values shown in Fig. \ref{seebeck_evo}. It is obtained that 
   $S_{\left( T\rightarrow \infty \right)}$ decreases in absolute value if doping level increases. It can be seen that $S_{\left( T\rightarrow \infty \right)}$ 
   is really strain dependent for low doping regimes while for the higher ones it becomes independent, which is in agreement with Fig. \ref{sto_deg}. Values for the effective masses $m^*/m_0$ obtained using eq. (\ref{effective_mass}) are also shown. Only high doping rates, i.e. greater than or equal to $2.0\%$, 
   can be considered relevant since it is where the free 
  electron gas approximation can be taken as a good model. Effective mass is larger for tensile strain.}
   \begin{minipage}[b]{80mm}
  \subcaption*{\textbf{Doping}  $\mathbf{0.1\%}$}
    \begin{tabular}{|c|c|c|c|c|}
    \hline
          & $A$ $(\mu V)$ & $S_{\left( T\rightarrow \infty \right)}$ $(\mu V/K)$ & $T_0$ $(K)$ & $m^*/m_0$\\
    \hline
  GSO & 163000 & -703 & 232 & 0.145\\
  DSO & 158000 & -730 & 217 & 0.150\\
  STO & 159000 & -737 & 215 & 0.151\\
  LSAT & 166000 & -730 & 228 & 0.145\\
  LAO & 174000 & -720 & 242 & 0.140\\
    \hline
    \end{tabular}%
    \bigskip    
    
    \subcaption*{\textbf{Doping} $\mathbf{1.0\%}$}
    \begin{tabular}{|c|c|c|c|c|}
    \hline
          & $A$ $(\mu V)$ & $S_{\left( T\rightarrow \infty \right)}$ $(\mu V/K)$ & $T_0$ $(K)$ & $m^*/m_0$\\
    \hline
GSO & 215000 & -533 & 403 & 0.509\\ 
DSO & 215000 & -537 & 400 & 0.513\\ 
STO & 219000 & -545 & 403 & 0.506\\
LSAT & 230000 & -543 & 423 & 0.489\\
LAO & 240000 & -534 & 450 & 0.473\\
    \hline
    \end{tabular}%
    
    \bigskip 
    \subcaption*{\textbf{Doping}  $\mathbf{4.0\%}$}
    \begin{tabular}{|c|c|c|c|c|}
    \hline
          & $A$ $(\mu V)$ & $S_{\left( T\rightarrow \infty \right)}$ $(\mu V/K)$ & $T_0$ $(K)$ & $m^*/m_0$\\
    \hline
GSO & 296000 & -425 & 696 & 0.932 \\
DSO & 303000 & -426 & 711 & 0.917 \\
STO & 311000 & -427 & 729 & 0.899 \\
LSAT & 325000 & -427 & 761 & 0.870 \\
LAO & 330000 & -425 & 775 & 0.868 \\
    \hline

    \end{tabular}%
    
    \end{minipage}
    \begin{minipage}[b]{80mm}

    \subcaption*{\textbf{Doping}  $\mathbf{0.5\%}$}  
    \begin{tabular}{|c|c|c|c|c|}
    \hline
          & $A$ $(\mu V)$ & $S_{\left( T\rightarrow \infty \right)}$ $(\mu V/K)$ & $T_0$ $(K)$ & $m^*/m_0$\\
    \hline
GSO & 195000 & -592 & 329 & 0.353\\ 
DSO & 199000 & -605 & 323 & 0.348\\
STO & 190000 & -609 & 312 & 0.368\\
LSAT & 199000 & -607 & 328 & 0.354\\ 
LAO & 210000 & -596 & 352 & 0.341\\
    \hline
    \end{tabular}%
    \bigskip
    \subcaption*{\textbf{Doping}  $\mathbf{2.0\%}$}
    \begin{tabular}{|c|c|c|c|c|}
    \hline
          & $A$ $(\mu V)$ & $S_{\left( T\rightarrow \infty \right)}$ $(\mu V/K)$ & $T_0$ $(K)$ & $m^*/m_0$\\
    \hline
GSO & 241000 & -481 & 501 & 0.720\\
DSO & 245000 & -483 & 508 & 0.714\\
STO & 252000 & -486 & 519 & 0.700\\
LSAT & 265000 & -485 & 546 & 0.672\\
LAO & 272000 & -481 & 567 & 0.661\\
    \hline
    \end{tabular}%
    
    \bigskip  
    \subcaption*{\textbf{Doping}  $\mathbf{50.0\%}$}
    \begin{tabular}{|c|c|c|c|c|}
    \hline
          & $A$ $(\mu V)$ & $S_{\left( T\rightarrow \infty \right)}$ $(\mu V/K)$ & $T_0$ $(K)$ & $m^*/m_0$\\
    \hline
GSO & 1490000 & -154 & 9690 & 0.995 \\ 
DSO & 1560000 & -154 & 10100 & 0.957 \\ 
STO & 1590000 & -154 & 10300 & 0.949 \\ 
LSAT & 1670000 & -154 & 10900 & 0.910 \\ 
LAO & 1750000 & -154 & 11400 & 0.878 \\ 
    \hline
    \end{tabular}%
    
    \end{minipage}
    
  \label{parameters}%
\end{table*}%

Table \ref{parameters} summarizes the results of the fit to our model (different parameters described in detail in the main text) for the thermopower at different doping levels and for the various strain situations considered.

\clearpage

\end{document}